\documentclass[12pt]{article}

\newcommand*{\seconds}{\mbox{ s}} 
\newcommand*{\cel}{{^{\circ}\mbox{ C}}}

\begin{document}

\title{Why are very short times
 so long and very long times so short in elastic waves?}

\author{Guido Parravicini and Serena Rigamonti
\\
Dipartimento di Fisica,
Universit\`a degli Studi di Milano, 
\\
via Celoria 16, I-20133
  Milano, Italy
\\
guido.parravicini@unimi.it} 

\maketitle

\begin{abstract}
In a first study of thermoelastic waves, such as on the textbook of
Landau and Lifshitz, one might at first glance understand that when
the given period is very short, waves are isentropic because heat
conduction does not set in, while if the given period is very long
waves are isothermal because there is enough time for thermalization
to be thoroughly accomplished. When one pursues the study of these
waves further, by the mathematical inspection of the complete
thermoelastic wave equation he finds that if the period is very short,
much shorter than a characteristic time of the material, the wave is
isothermal, while if it is very long, much longer than the
characteristic time, the wave is isentropic. One also learns that this
fact is supported by experiments: at low frequencies the elastic waves
are isentropic, while they are isothermal when the frequencies are so
high that can be attained in few cases. The authors show that there is
no contradiction between the first glance understanding and the
mathematical treatment of the elastic wave equation: for thermal
effects very long periods are so short and very short periods are so
long.
\end{abstract}


\section{Introduction}
In the deduction of the wave equation from the constitutive equations
of an elastic medium where thermal effects are taken into account, or
thermoelastic medium, one encounters the problem of choosing the
thermodynamical transformation undergone by the system during a
period. The alternative is between isothermal and adiabatic
transformations, this second case being isentropic because the
transformations are ideally considered reversible; the same problem
appears in the study of acoustic waves in fluids and is dealt with the
same arguments. The choice can be experimentally tested because the
velocity of the longitudinal waves in solid media and in fluids is
different in the isothermal and isentropic cases.

There might arise some confusion when the range of validity for the
two situations is stated on the basis of physical considerations only,
without the full mathematical derivation.  For example, in the study
of the chapter devoted to waves in the book by Landau and Lifshitz on
elasticity \cite{landau2}, one finds the statement: ``\textit{If the
heat exchange during times of the order of the period of oscillatory
motions in the body is negligible, we can regard any part of the body
as thermally insulated, i.~e.\ the motion is adiabatic.}''
\\
Students and physicists who do not pursue the study of elastic waves
further usually read this sentence as meaning that the motion is
adiabatic if the period is very short and that it is isothermal if the
period is very long.  One would therefore expect that waves with low
frequencies are isothermal, waves with high frequencies are
isentropic.
\\
Those who study elastic waves further find that mathematical arguments
\cite{chadwick, bowen} show that in order for a wave to be isentropic
its frequency has to be low with respect to a ``characteristic
frequency'' or, otherwise stated, that the period must be long enough,
much longer than a ``characteristic time'' of the material to be
defined below in (\ref{eq tau characteristic}).  This is supported by
experiments.  At the same time, mathematics shows that waves are
isothermal when their frequency is very high with respect to the
characteristic frequency, or when the period is very short compared to
the characteristic time \cite{chadwick, bowen}. We shall spend some
more words on these facts in section~\ref{sec statement}.

We have an apparent contradiction between a compelling mathematical
fact, supported by experiments, and the elementary physical argument
exemplified by Landau and Lifshitz' textbook; because of this
contradiction, the latter should be discarded.
\\
This is not the case.  In section~\ref{sec landau}, we show that the
elementary physical argument and the mathematical one do not conflict
with each other because both read as: \emph{when the period is so short
that heat conduction does not set in, that is to say when the period
is much longer than the characteristic time, the wave is adiabatic;
when the period is so long that thermalization occurs, that is to say
when the period is much shorter than the characteristic time, the wave
is isothermal.}
\\
In other words, very short periods are so long and very long periods
are so short.

The core of our argument is that when we say that the period is short,
 or long, we must first declare which time we are comparing the period
 to, as we should never forget to do when stating that a quantity is
 small or large.  The ``comparison time'' suited to thermoelastic
 waves turns out to be proportional to the square of the period
 itself; it follows that the ratio between the period and the
 comparison time is inversely proportional to the period. The constant
 of proportionality is the characteristic time.  Therefore, when the
 ratio of the period into the comparison time is very large, the
 period is very short compared to the characteristic time so that the
 wave is isothermal according to both the elementary physical
 considerations and mathematics. On the other hand, when the ratio
 between the period and the comparison time is very small, the period
 is much longer than the characteristic time, and the wave is
 isentropic according to the physical considerations, to mathematics,
 and to experiments.

We limit our argument to one-dimensional thermoelastic waves as its
extension to the two- and three-dimensional cases is trivial. The same
line of reasoning followed here holds true for acoustic waves in
fluids as well.

This article is intended for both the undergraduate students who start
the study of acoustics beyond the elementary level and the graduate
students who do not specialize in acoustics.

\section{
The Characteristic Time and Frequency 
 \label{sec statement}}

The mathematical deduction of the wave equation leads to the
conclusion that waves are adiabatic or isothermal according to whether
the period is much longer or much shorter than the characteristic time
of the material.  We devote some time to this point, following
\cite{bowen} and its notation, even though it might be slightly
unconventional.
\\
 The thermoelastic wave equation, see  (1.11.16) of
 \cite{bowen}, reads:
\begin{equation}
\frac{\partial}{\partial t}\left(\frac{\partial^2 w}{\partial t^2} -
     {a^*}^2 \frac{\partial^2 w}{\partial X^2} \right) -a^2
\tau_{\kappa} \frac {\partial^2}{\partial X^2}\left(\frac{\partial^2
     w}{\partial t^2} - 
     a^2 \frac{\partial^2 w}{\partial X^2} \right) =0\,. 
\label{eq bowen complete}
\end{equation}
 Here, $w=w(X,t)$ can be either the displacement of an element of the
medium or its temperature, $a^*$ and $a$ are respectively the
isentropic and the isothermal velocity, and $\tau_{\kappa}$ is a
characteristic time of the material defined by the relation:
\begin{equation}
\tau_{\kappa}=\frac{\kappa}{\rho c_v a^2}\,,   \label{eq tau characteristic}
\end{equation}
where $\kappa$ is the coefficient of thermal conduction, $\rho$ the
mass density, and $c_v$ the specific heat at constant volume. The fact
that the characteristic time depends upon the properties of the
material only is quite useful and important.  
\\ 
Inspection of (\ref{eq bowen complete}) shows that if the wave period
$T$ is much shorter than $\tau_{\kappa}$, the first term is negligible
in front of the second one so that the quantity $w$ propagates as an
isothermal wave.  Instead, when the period is much longer than
$\tau_{\kappa}$, the second term can be disregarded so that the
quantity $w$ satisfies the equation of the isentropic wave.  
\\ 
This is of course equivalent to stating that if the frequency is much
higher than the characteristic frequency $1/\tau_{\kappa}$ the wave is
isothermal, whereas it is isentropic when its frequency is much lower
than the characteristic frequency. Usually, the argument is stated in
terms of frequency rather than of time.  
\\ 
The characteristic times of four metals at temperature $20 \cel$
are given in Table 1.11.1 of \cite{bowen}, after \cite{chadwick};
their order of magnitude is $10^{-12}\seconds$. Further on in the same
section 1.11 of \cite{bowen}, the characteristic time for air at
standard pressure and $20 \cel$ is computed, and its order of
magnitude turns out to be $10^{-10}\seconds$.  
\\ 
Within the nowadays achievable range of frequencies, experiments
confirm the mathematical statement for the isentropic case, e.~g.\ see
\cite{chadwick}.  Tests for the isothermal case are more difficult
instead: in elastic media the frequencies necessary for the isothermal
wave are above the terahertz, which is the upper limit attained today
in the production of acoustic waves in solids, for example see
\cite{saser}; experiments that exhibit the transition of the sound
velocity from its adiabatic to its isothermal value in liquid metals
are rather recent, as an example see \cite{metalli liquidi}.

\section{Very short times are so long, very long times are so short
 \label{sec landau}} 

We consider monochromatic waves with period $T$ and wavelength
$\lambda$.  We shall also use ``wavelength'' for short in place of ``a
part of the continuum as long as a wavelength.'' By ``thermalization
time'' of a part of a continuum, we mean the time that has to elapse
so that the temperature of that part of the continuum equalizes, or a
disturbance of the temperature be smoothed away.

The comparison time referred to in the introduction is the
thermalization time of a wavelength, to be determined further on,
because a monochromatic wave is periodic in space so that what happens
to a single such wavelength happens to the whole wave.

Let $\tau(\lambda)$ be the thermalization time of the wavelength
$\lambda$.  We rephrase the elementary physical argument as follows.
\begin{enumerate}
\item \label{item isentropic} \emph{When the period of the wave is
very short compared to the time of thermalization of a wavelength,
heat conduction cannot set in so that the wave is adiabatic by
definition.}  If the inequality
\begin{equation}
T\ll \tau(\lambda) \label{eq very short}
\end{equation}
holds the wave is adiabatic and therefore isentropic.
\item \label{item isothermal} \emph{When the period of the wave is
very long compared to the thermalization time of a wavelength, the
wave is isothermal by the zeroth law of thermodynamics, because heat
conduction does set in and there is enough time for thermalization.}
If the inequality
\begin{equation}
T\gg \tau(\lambda) \label{eq very long}
\end{equation}
holds the wave is isothermal.
\end{enumerate}
Next, we compute $\tau(\lambda)$. As long as we are interested not in
exact values but in orders of magnitude, we are free to evaluate the
thermalization time from the heat kernel of the mechanically static
heat conduction which is a combination of exponential functions of the
form $\exp -X^2/4 \chi t$, where the particular combination depends
upon the boundary conditions. The coefficient $\chi$ is the thermal
diffusion coefficient defined by the relation:
\begin{equation}
  \chi = \frac{\kappa}
{\rho c_v} \,.
\label{eq chi} 
\end{equation}
The thermalization time $\tau(\lambda)$ for a piece of material of
length $\lambda$ is such that the argument of the exponential of the
heat kernel be about equal to minus one: $\lambda^2/\chi \tau(\lambda)
\sim 1$, for example see \cite{landau1}.  Therefore, the approximate
equalities hold:
\begin{equation}
  \tau(\lambda) \sim \frac{\lambda^2}{\chi} \sim \frac{{a^*}^2 T^2}{ \chi}\,.
\label{eq thermalization time}  
\end{equation}
Whether we use the isentropic velocity $a^*$ or the isothermal one $a$
in the relation $\lambda=aT$, is of no consequence for (\ref{eq
thermalization time}) as we are dealing with orders of magnitude only.
Because of the definitions, (\ref{eq tau characteristic}) and (\ref{eq
chi}), of the characteristic time $\tau_{\kappa}$ and of the thermal
diffusion coefficient $\chi$, the approximate equality~(\ref{eq
thermalization time}) becomes:
\begin{equation}
  \tau(\lambda) \sim \frac{T^2}{\tau_{\kappa}}\,.
\label{eq thermalization and characteristic time}  
\end{equation}
This is the relation we mentioned in the introduction about the
comparison time and the period of the wave. Incidentally, this formula
clarifies the physical significance of the characteristic time
$\tau_{\kappa}$, wich is the period $T_{\kappa}$ of that particular
wavelength $\lambda_{\kappa}$ whose thermalization time
$\tau(\lambda_{\kappa})$ is the same as the period $T_{\kappa}$:
\begin{equation}
  \tau_{\kappa}=T_{\kappa}= \tau(\lambda_{\kappa})\,. 
\label{eq meaning of characteristic tau}
\end{equation}
In order for the wave to be isentropic according to the elementary
physical argument, we use (\ref{eq very short}) and~(\ref{eq
thermalization and characteristic time}), and obtain the following
relation:
\begin{equation}
 T \ll \tau(\lambda) \Leftrightarrow T \ll  \frac{T^2}{\tau_{\kappa}}\,.
\end{equation}
Therefore, the  inequalities  
\begin{equation}
 T \ll \tau(\lambda)\> \mbox{ and }\> T\gg \tau_{\kappa} 
\label{eq short then long}
\end{equation}
are equivalent so that when the period $T$ is very short in front of
the thermalization time $\tau(\lambda)$ of the wavelength, it is very
long in front of the characteristic time, and the wave is isentropic
according to mathematics as well.  The same holds for the isothermal
case by exchanging short for long and long for short, and reversing
the signs of the inequalities in (\ref{eq short then long}).
\\
One might object that the phenomenon under study is definitely not
mechanically static so that we may not use the heat kernel that led us
to the evaluation of the thermalization time given in (\ref{eq
thermalization time}) and~(\ref{eq thermalization and characteristic
time}).  However, the thermalization time is explicitly computed in
\cite{chadwick}, p.~293:
\begin{equation}
  \tau(\lambda)= \frac{\rho c_v (1+\beta )}{\kappa} \left(\frac{
    \lambda}{2\pi} \right)^2 \,. 
\label{eq tempo di decadimento di chadwick}
\end{equation}
The dimensionless coefficient $\beta$ is the ``decay parameter''; its
order of magnitude for the metals considered in \cite{chadwick} is
$10^{-2}$.  From this formula, one draws (\ref{eq thermalization
time}), (\ref{eq thermalization and characteristic time}), and then
(\ref{eq short then long}) and its counterpart for the isothermal wave
as before.

Therefore, the elementary physical argument is vindicated: it is the
same as the mathematical argument.


\end{document}